\documentclass{kluwer}
\usepackage{graphicx} 
\begin{document}
\begin{article}
\begin{opening}
\title{Optimization code with weighting function
for the reconstruction of  coronal magnetic fields.}

\author{T. \surname{Wiegelmann}
\email{wiegelmann@linmpi.mpg.de}}
\institute{Max-Planck-Institut f\"ur Aeronomie,
Max-Planck-Strasse 2, 37191 Katlenburg-Lindau, Germany}

\date{DOI: 10.1023/B:SOLA.0000021799.39465.36 \\
Bibliographic Code: 2004SoPh..219...87W}



\runningtitle{Optimization code}
\runningauthor{Wiegelmann}

\begin{ao}
Kluwer Prepress Department\\
P.O. Box 990\\
3300 AZ Dordrecht\\
The Netherlands
\end{ao}

\begin{motto}

\end{motto}

\begin{abstract}
We developed  a code for the reconstruction of nonlinear force-free and
not force-free coronal magnetic fields. The 3D
magnetic field is computed numerically with the help of an optimization
principle. The force-free and non force-free codes are compiled in one
program. The force-free approach needs photospheric vectormagnetograms
as input. The non force-free code additional requires the line-of-sight
integrated coronal density distribution in
combination with a tomographic inversion code.
Previously the optimization approach has been used to compute magnetic
fields using all six  boundaries of a computational box. Here we extend
this method and show how the coronal magnetic field can be reconstructed only
from the bottom boundary, where the boundary conditions are measured with
vector magnetographs.
The program is planed for use within the Stereo mission.
\end{abstract}


\abbreviations{\abbrev{KAP}{Kluwer Academic Publishers};
   \abbrev{compuscript}{Electronically submitted article}}

\nomenclature{\nomen{KAP}{Kluwer Academic Publishers};
   \nomen{compuscript}{Electronically submitted article}}

\classification{JEL codes}{D24, L60, 047}
\end{opening}
\section{Introduction}
\label{introduction}
The solar magnetic field is an important quantity which couples the solar
interior with the photosphere and atmosphere. Knowledge regarding the coronal
magnetic field plays a key role for eruptive phenomena, e.g. coronal mass
ejection, flares and eruptive prominences. Unfortunately a direct measurement
of the coronal magnetic field is extremely difficult. In principle one can
use the polarization of emissions from magnetic sensitive coronal line transitions
to draw conclusions about the coronal magnetic field. These lines,
however, are very faint so that in the past they have only occasionally been
observed (e.g., \cite{house77,arnaud87,judge98}).
In a recent study \cite{judge01} conclude that several forbidden lines
(e.g. in Fe XIII, He I, Mg VIII and Si IX)  may be used to
determine the coronal magnetic field. They further
concluded that space born missions are not needed  for such kind of
coronal magnetometers but a high, dry mountain site. In their study they propose
a focal plane instrument devoted to the $1 \mu m$ region. These authors
also point out that besides  the observational part, a further major
problem is the interpretation of the data.
The line-of-sight integration inherent in these observations
make the data analysis a badly posed inversion problem.
Presently, algorithms based on vector tomography are studied to find out to which
extend the coronal magnetic field can be reconstructed from these
observations (Maxim Kramar and Bernd Inhester, private communication).

Despite these promising new developments regarding the
principle possibility of coronal B-field measurements we have to face the fact
that currently and probably also in near future high quality direct measurements
of the coronal magnetic field are not available.
As an alternative to the above measurements, a number of authors have
modelled the coronal magnetic field by extrapolation from more sound
photospheric magnetic field observations.

It is generally assumed that the magnetic pressure in the corona is much
higher than the plasma pressure (small plasma $\beta$) and that
therefore the magnetic field is nearly force-free (for a critical
view of this assumption see \cite{gary01}). The extrapolation methods
based on this assumption include potential field extrapolation \cite{schmidt64,semel67},
linear force-free field extrapolation \cite{chiu77,seehafer78,seehafer82,semel88} and
nonlinear force-free field extrapolation \cite{amari97}.
Methods for the extrapolation  non force-free fields and have been developed
by \cite{petrie00,twbi03}.
Potential fields can be determined directly from
line-of-sight magnetogram data (e.g. MDI on SOHO). Linear
force-free fields can as well be calculated from line-of-sight
magnetograms, but contain a free parameter $\alpha$ which has to be computed
from additional data, e.g. with fitting procedures which try to match the
field model with observed coronal plasma
loops using data from e.g. EIT \cite{twtn02} or Yohkoh \cite{laura03}.
Unfortunately potential
fields and linear force-free fields do not contain free energy and are very
probably a poor approximation for an active region prior to an eruption.
By free energy we understand energy which can be released during an
eruption. A linear force free field has more energy than a potential field. This energy
can, however, not be released during an eruption related to ideal or resistive
MHD-instabilities because a linear force free field cannot rapidly relax to a potential
field.  The reason is that the magnetic helicity is strictly  conserved
for ideal MHD and approximately conserved for resistive processes.
(The magnetic helicity is dissipated slower than the magnetic energy, see \cite{berger84}).
A nonlinear force free field can, however, relax to a linear force free field with the
same magnetic helicity. In this sense a nonlinear force free field has free
energy available for an eruption. Consequently investigations
regarding non-linear force-free fields are essential to understand eruptive
phenomena.

The calculation of non linear force-free fields is complicated by
the intrinsic nonlinearity of the underlying mathematical problem. From the
observational point of view the non linear reconstruction is also more
challenging because photospheric vector magnetograph data are required.
Unfortunately the transversal component of the photospheric B-field is
measured with significant lower accuracy as the line of sight component.
An additional problem is that the transversal magnetic field is only known
with respect to an $180^o$ ambiguity and a preprocessing of the raw data is
necessary to resolve this ambiguity (One possibility is the minimum energy method by
\cite{metcalf94} used for vector magnetogram data from IVM in Hawaii.).

Several methods have been proposed to compute nonlinear force-free
fields:
\begin{itemize}
\item A conceptionally simple method is to reformulate the force-free equations
(\ref{ampere-ff}-\ref{solenoidal-ff}) in such way that they can be used for an upward
integration of the vector magnetogram into the corona \cite{wu90,amari97}.
This direct extrapolation is an ill posed problem for the elliptic
equations (\ref{ampere-ff}-\ref{solenoidal-ff}) and consequently the method is
limited to low heights. In particular one finds that an erroneous
exponential growth of the magnetic field with increasing height is a typical
behaviour.
\item An alternative approach is to use a Grad-Rubin method
(\cite{sakurai81,amari99}).
This method uses a potential field as initial equilibrium and then
progressively currents are introduced into the system and the fields
are relaxed towards a force-free state. The method is especial useful
for small deviations from a potential field with small values of $\alpha$
and modest non linearities. The method requires an explicit calculation of
the $\alpha$ distribution on the photosphere. In principle the computation
of $\alpha$ is straight forward  $\alpha(x,y)=\frac{\partial B_y/\partial x-\partial B_x/\partial
y}{B_z(x,y)}$ but this is inaccurate for
observational data for the following reasons. First one needs the transversal
component of the photospheric magnetic field which is measured with lower
accuracy than the line of sight magnetic field. In a second step one has to
take the horizontal derivatives (x,y)of these inaccurate values and finally one
has to divide through the normal magnetic field $B_z$  which causes additional
problems where $|B_z|$ is small. These errors cumulate in the photospheric $\alpha$
distribution.
\item A third possibility is to use
the method of MHD relaxation \cite{chodura81,roumeliotis96}. The idea is to start
with a suitable magnetic field which is not in equilibrium
and to relax it into a force-free state. For test configurations \cite{lowandlou} the MHD
relaxation method
converges to the exact solution, but with less accuracy than the optimization
approach discussed below \cite{twtn03}.
\item In the optimization approach \cite{wheatland00} a functional
containing the force-free equations is minimized. The method directly
uses the measured vector magnetograph data and an explicit computation of $\alpha$ is
not necessary. Another advantage of the method is that the quality of the reconstructed
magnetic field (force-free and solenoidal condition) is controlled automatically within the
iteration procedure. A difficulty of the method is that it requires boundary conditions
on all boundaries of a computational box while for observational data only the bottom
boundary data are known. Within this paper we are dealing with this problem and extend
the optimization method accordingly.
\end{itemize}
While the low $\beta$ plasma in the lower corona can be described with the non linear
force-free approach, it is necessary to include plasma pressure and solar
gravitation to describe regions with a finite plasma $\beta$, e.g. helmet streamers.
\cite{twbi03} extended the force-free optimization with the aim to include these forces
and showed that the method converges for test configurations. The
difficulties regarding the lateral and top boundaries are analog to the non-linear
force-free case. The non
force-free reconstruction requires additional data as input, e.g. the coronal plasma
density distribution from an assumed model or computed with help of tomographic methods.

We outline the paper as follows. In section \ref{basic} we provide the basic equations
of the modified optimization method
and derive the iteration equations. In section \ref{special} we specify useful forms of
the weighting function in the boundary regions. Section \ref{test1} contains test runs
regarding the non linear force-free case and  section \ref{test2} consistency
checks for non force-free configurations. We draw conclusions in
section \ref{conclusions} and give an outlook for further research.
\section{Basic equations}
\label{basic}
Force-free coronal magnetic fields have to obey the equations
\begin{eqnarray}
{\bf j}\times{\bf B} & = & {\bf 0},
\label{forcefree}\\
\nabla \times {\bf B }& = & \mu_0 {\bf j}  \label{ampere-ff}, \\
\nabla\cdot{\bf B}    & = &         0      \label{solenoidal-ff}.
\end{eqnarray}
The force-free approach is valid in the low corona where the plasma $\beta$ is
small. For extended structures, e.g. helmet streamers the plasma $\beta$
increases and the force-free assumption is not valid anymore. Therefore it
is necessary to consider the effect of plasma pressure and gravity here and
solve the magneto hydro static equations (MHS).
\begin{eqnarray}
{\bf j}\times{\bf B} -\nabla P -\rho \nabla \Psi  & = & {\bf 0},
\label{forcebal}\\
\nabla \times {\bf B }& = & \mu_0 {\bf j}  \label{ampere}, \\
\nabla\cdot{\bf B}    & = &         0      \label{solenoidal},
\end{eqnarray}
where ${\bf B}$ is the magnetic field, ${\bf j}$ the electric current
density, $P$ the plasma pressure, $\rho$ the plasma density,
$\mu_0$ the vacuum permeability and $\Psi$ the solar gravity potential.
We define the functional
\begin{equation}
L=\int_{V} \; w(x,y,z) \; B^2 \; (\Omega_a^2+\Omega_b^2) \; d^3x
\label{defL},
\end{equation}
with

\begin{eqnarray}
{\bf \Omega_a} &=& \; \left \{
\begin{array}{ll}
B^{-2} \;\left[(\nabla \times {\bf B})
  \right] & (\mbox{force-free fields}) \\
          B^{-2} \;\left[(\nabla \times {\bf B})
\times {\bf B} - \mu_0 (\nabla P +\rho \nabla \Psi) \right]
& (\mbox{MHS})
\end{array}
\right. \\
{\bf \Omega_b} &=& B^{-2} \;\left[(\nabla \cdot {\bf B}) \; {\bf B} \right].
\label{defomega}
\end{eqnarray}
$w(x,y,z)$ is a weighting function. Useful forms of the weighting function
will be discussed below.

For the force-free case the functional is given explicitly as
\begin{equation}
L=\int_{V} \; w(x,y,z) \; \left[B^{-2} \, |(\nabla \times {\bf B}) \times {\bf B}|^2
+|\nabla \cdot {\bf B}|^2\right] \; d^3x
\label{defL1},
\end{equation}
and it is obvious that (for $w>0$) the force-free equations
(\ref{forcefree}-\ref{solenoidal-ff}) are fulfilled when L is equal zero.
For the non force-free case L is given explicitly as
\begin{equation}
L=\int_{V} \; w(x,y,z) \; \left[B^{-2} \, |(\nabla \times {\bf B}) \times {\bf B} -\mu_0
(\nabla P +\rho \nabla \Psi)|^2 +|\nabla \cdot {\bf B}|^2\right] \; d^3x
\label{defL2},
\end{equation}
and when the functional reaches (for $w>0$) its minimum at $L=0$ then the MHS equations
(\ref{forcebal}-\ref{solenoidal}) are fulfilled.

The following
discussion is equivalent for the force-free and non force-free case.
Without weighting function $(w=1)$ the method has been developed by \cite{wheatland00}
for the force-free case and by \cite{twbi03} for the non force-free case.
For $w(x,y,z)=1$ the optimization method requires that the magnetic
field is given on all (6 for a rectangular computational box) boundaries.
This causes a serious limitation of the method because such data are only
available for model configurations.
For the reconstruction of the coronal magnetic field it is necessary to develop a
method which reconstructs the magnetic field only from photospheric vector magnetograms. Vector
magnetograms provide boundary conditions only for the bottom boundary  of a computational
box while the other five boundaries remain unknown. Without a weighting function all
six boundaries of the
computational box have equal rights and influence the solution in the box.
It is therefore important to diminish the effect
of the top and lateral boundaries on the magnetic field inside the computational box. This can be
done either by including a variation of B not only in the interior but also
on those boundaries where B is unknown. This approach, however, is
numerically difficult because it involves two types of variations. We show
that it is essentially equivalent to introducing finite size boundary regions
on those boundaries where B is unknown with
the weighting function  $w(x,y,z)$ different from unity.

The idea is do define an interior physical region
where we want to calculate the magnetic
field so that it fulfills the force-free or MHS equations.
This region is in the center of the box (including the photosphere) with $w=1$. The computational
box additionally includes boundary layers towards
the lateral and top boundary where $w$ decreases to $0$ at the computational boundary.
Consequently the method weights deviations from the force-free state (or MHS-state) less
severely close to the boundary.
The use of a weighting
function has been proposed for the force-free case by \cite{wheatland00} in the conclusions but
no iteration equations or test simulations have been presented.
Here we provide these iteration equations for the
more generalized case. We carry out several tests to investigate the optimum shape
of the weighting function and how the size of the boundary layer influences the quality of
the reconstruction.

We minimize equation (\ref{defL}) with respect to an iteration parameter $t$
(see Appendix \ref{appendixA} for details) and obtain an iteration equation for
the magnetic field
\begin{equation}
\Rightarrow \frac{1}{2} \; \frac{d L}{d t}=-\int_{V} \frac{\partial {\bf B}}{\partial t} \cdot {\bf \tilde{F}} \; d^3x
-\int_{S} \frac{\partial {\bf B}}{\partial t} \cdot {\bf \tilde{G}} \; d^2x
\label{minimize1}
\end{equation}
\begin{eqnarray}
{\bf \tilde{F}}&=& w \; {\bf F} +({\bf \Omega_a} \times {\bf B}
)\times \nabla w  +({\bf \Omega_b} \cdot {\bf B}) \; \nabla w \\
{\bf \tilde{G}}&=& w \; {\bf G}
\end{eqnarray}
\begin{eqnarray}
{\bf F} & =& \nabla \times ({\bf \Omega_a} \times {\bf B} )
- \bf \Omega_a \times (\nabla \times \bf B)  \nonumber\\
& & +\nabla(\bf \Omega_b \cdot \bf B)-  \bf \Omega_b(\nabla \cdot \bf B)
+( \Omega_a^2 + \Omega_b^2)\; \bf B
\end{eqnarray}
\begin{eqnarray}
{\bf G} & = & {\bf \hat n} \times ({\bf \Omega_a} \times {\bf B} )
-{\bf \hat n} (\bf \Omega_b \cdot \bf B),
\label{defG}
\end{eqnarray}
and $\hat n$ is the inward unit vector on the surface $S$.
The surface integral in (\ref{minimize1}) vanishes if the magnetic field is described on the
boundaries of a computational box. Inside the computational box we iterate
the magnetic field with
\begin{equation}
\frac{\partial {\bf B}}{\partial t} =\mu {\bf \tilde{F}}
\label{iterateB},
\end{equation}
which insures that $L$ is monotonically decreasing.
\subsection{Algorithm}
\label{Algorithm}
We compute the 3D-coronal magnetic field in a numerical box using the
following steps.
\begin{itemize}
\item As a start configuration we use the measured normal component $B_z$
of the magnetic field to calculate a potential magnetic field in the
whole box with help of a Fourier representation \cite{seehafer78}.
\item For non force-free (finite $\beta$) configurations the plasma density distribution
is described in the box. This step is unnecessary for force-free $(\beta \ll 1)$
fields.
\item We use vector magnetograph data to describe the bottom boundary
(photosphere) of the computational box. On the lateral and top boundaries the
field is chosen from the potential field above.
\item We iterate for the magnetic field inside the computational box with
(\ref{iterateB}) using a Landweber-iteration (see e.g. \cite{louis}).
The continuous form of (\ref{iterateB}) guaranties a monotonically decreasing
$L$. This is as well ensured  in the discretized form  if
the iteration step $dt$ is sufficiently small.
The code checks if
$L(t+dt) < L(t)$ after each time step. If the condition is not fulfilled,
the iteration step is repeated with $dt$ reduced by a factor of 2.
After each successful iteration step we increase $dt$ slowly by a factor
of $1.01$ to allow the time step to become as large as possible with respect
to the stability condition. The iteration stops if $dt$ falls below a limiting
value, e.g. $1/100$ of the initial iteration step
\footnote{We find that the time step $dt$ keeps on decreasing recurrently when
the solution has converged. We never found a further improvement of L after
$dt$ has  once fallen below $1/100$ of the initial iteration step.}
in the current version of the code.
\end{itemize}
Let us remark that the main numerics of the optimization code is similar for the method
with and without a weighting function. The main problem for the optimization
method without weighting function is that it requires the vector magnetic
field on all six boundaries of a computational box. As only the bottom
boundary is measured one has to make assumptions regarding the lateral and
top boundary, e.g. assume a potential field. In general this leads to
inconsistent boundary conditions (See \cite{aly89} regarding the compatibility of
photospheric vector magnetograph data.) and consequently a bad quality of the
reconstructed magnetic field. With help of the weighting function the five
inconsistent boundaries are replaced by boundary layers and consequently
we get more flexible boundaries around the physical domain which will adjust automatically during
the iteration. The idea of introducing a boundary layer with $w<1$ is to
reduce the dependence of the solution in the interior of the box from the unknown boundary
conditions. Since we have no measurements on these boundaries any choice of the
boundary conditions is a mere guess. The aim is only to
allow the solution in the interior to evolve more independently from the
boundary conditions chosen. So the advantage of the boundary layer is a
higher degree of
independence of the solution in the interior from the chosen boundary.
The price we have to pay is a higher computing time, as the
magnetic field has to be iterated within the whole computational box which
includes the physical domain as well as the boundary layers.
\section{Special forms of the weighting function $w(x,y,z)$.}
\label{special}
We want to use the weighting function to deal with
the unknown top and lateral boundaries. We define an inner {\it physical}
domain $V_i$ with $w=1$ and boundary layers $V_b$ where $w$ decreases monotonically from
1 to 0 through  the outer {\it numerical} boundary layer with the thickness $d$.
Consequently $w$ becomes
one-dimensional in each boundary layer (e.g. $w=w(z)$ at the top boundary
layer) and we get $\nabla w= \hat n \; \frac{\partial w}{\partial \hat n}$.
The surface integrals vanish on all boundaries because $w=0$ on the top and
lateral boundaries and $\frac{\partial {\bf B}}{\partial t}=0$ on the bottom
boundary where the magnetic field is measured with vectormagnetographs.
Consequently (\ref{minimize1}) reduces to
\begin{eqnarray}
\frac{1}{2} \; \frac{d L}{d t}&=&-\int_{V_i} \frac{\partial {\bf B}}{\partial t} \cdot {\bf  F} \; d^3x
-\int_{V_b} \frac{\partial {\bf B}}{\partial t} \cdot {\bf  \tilde{F}} \; d^3x
\\
\Rightarrow \frac{1}{2} \; \frac{d L}{d t}&=&-\int_{V_i} \frac{\partial {\bf B}}{\partial t} \cdot {\bf  F} \; d^3x
-\int_{V_b} \; w \; \frac{\partial {\bf B}}{\partial t} \cdot {\bf F} \; d^3x
\nonumber \\
&& -\int_{V_b} \frac{\partial {\bf B}}{\partial t} \cdot \left[ ({\bf \Omega_a} \times {\bf B}
)\times \nabla w \;   +({\bf \Omega_b} \cdot {\bf B}) \; \nabla w \right]   \; d^3x \\
\Rightarrow \frac{1}{2} \; \frac{d L}{d t}&=&-\int_{V_i} \frac{\partial {\bf B}}{\partial t} \cdot {\bf  F} \; d^3x
-\int_{V_b} \; w \; \frac{\partial {\bf B}}{\partial t} \cdot {\bf F} \; d^3x
\nonumber \\
&& -\int_{V_b} \frac{\partial w}{\partial \hat n}\;
\frac{\partial {\bf B}}{\partial t} \cdot \left[ \hat n \; \times ({\bf \Omega_a} \times {\bf B})
- \hat n \; ({\bf \Omega_b} \cdot {\bf B}) \right] d^3x
\label{minimize2}.
\end{eqnarray}
It is interesting to investigate the limit of an infinitesimally thin boundary
layer $d \rightarrow 0$ in (\ref{minimize2}). The thinner the boundary layer
becomes the steeper is $\frac{\partial w}{\partial \hat n}$ and for an
infinitesimally thin boundary layer the gradient becomes infinity.
The boundary layer is constructed in such way that independent
from the sheet thickness $d$ we have
$\int_0^d \frac{\partial w}{\partial \hat n}\;d \hat n=1$
which remains true also for $d \rightarrow 0$. In the limit
of $d \rightarrow 0$ the term $\hat n \; \times ({\bf \Omega_a} \times {\bf B})
- \hat n \; ({\bf \Omega_b} \cdot {\bf B})$ remains constant through the
sheet and the integration regarding $d\hat n$ can be carried out
explicitly. Consequently only a surface integral remains as the last term
in equation (\ref{minimize2}). The second integral in (\ref{minimize2})
vanishes for $d \rightarrow 0$ and we get
\begin{equation}
\frac{1}{2} \; \frac{d L}{d t}=-\int_{V} \frac{\partial {\bf B}}{\partial t} \cdot {\bf F} \; d^3x
-\int_{S} \frac{\partial {\bf B}}{\partial t} \cdot {\bf G} \; d^2x.
\label{minimize3}
\end{equation}
This exactly coincides with the non weighted case. Consequently equation (\ref{minimize2})
is a generalization of the usual optimization equation (\ref{minimize3}).
\begin{figure}
\hspace*{\fill}
\includegraphics[clip,height=6.0cm]{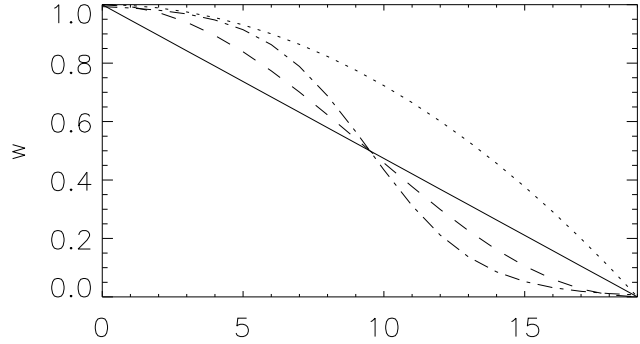}
\hspace*{\fill}
\caption{Different profiles for the weighting function $w$. The solid lines shows a linear profile,
the dotted line a quadratic, the dashed line a cos and the dashdotted line a
tanh profile. All profiles are equal one at the physical boundary, decrease
monotonically within the boundary layer and reach zero at the boundary of the computational box.}
\label{wprofil}
\end{figure}
\section{Tests for non linear force-free configurations.}
\label{test1}
\begin{figure}
\hspace*{\fill}
\mbox{
\includegraphics[clip,height=6.0cm]{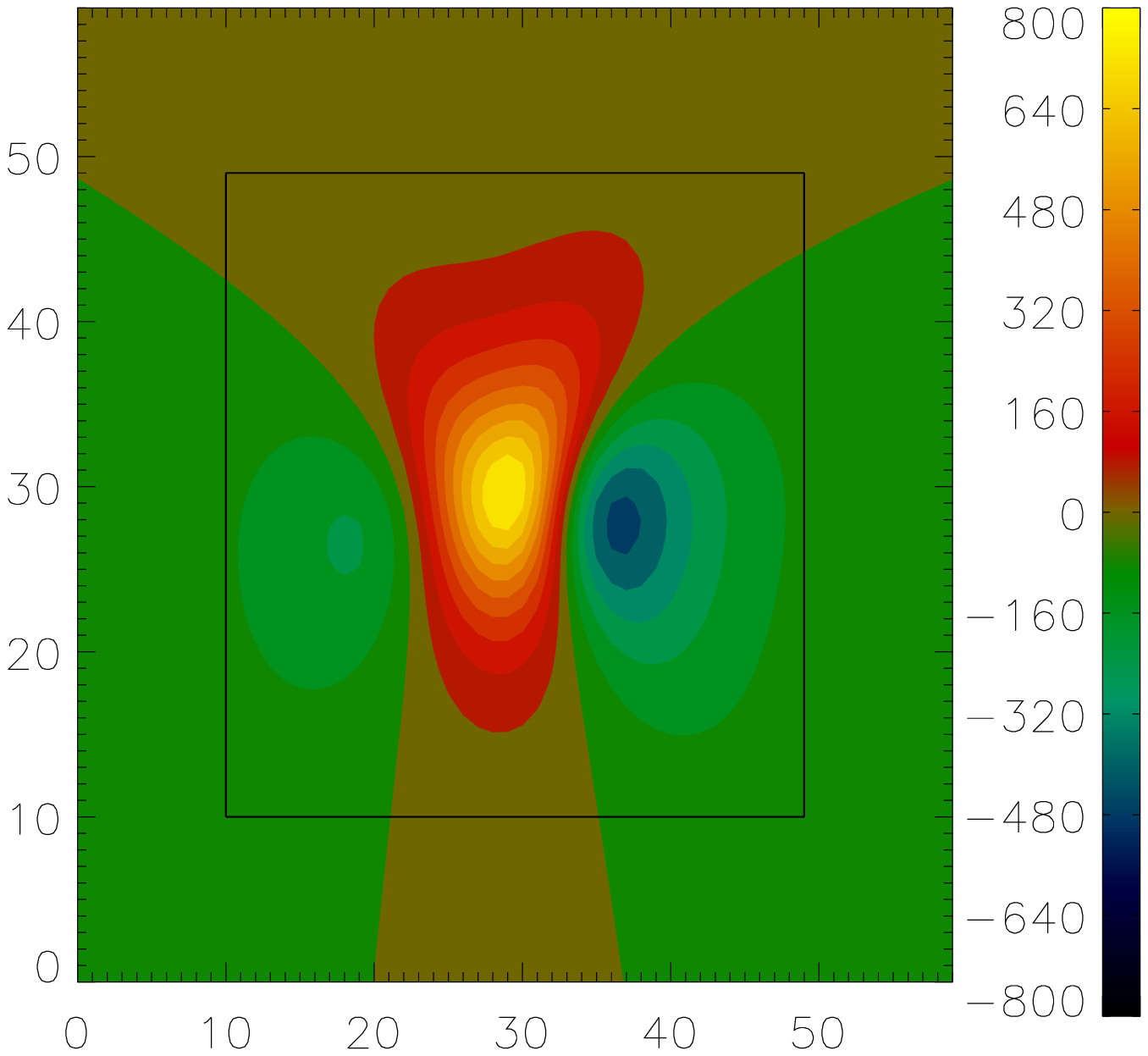}
\includegraphics[clip,height=6.0cm]{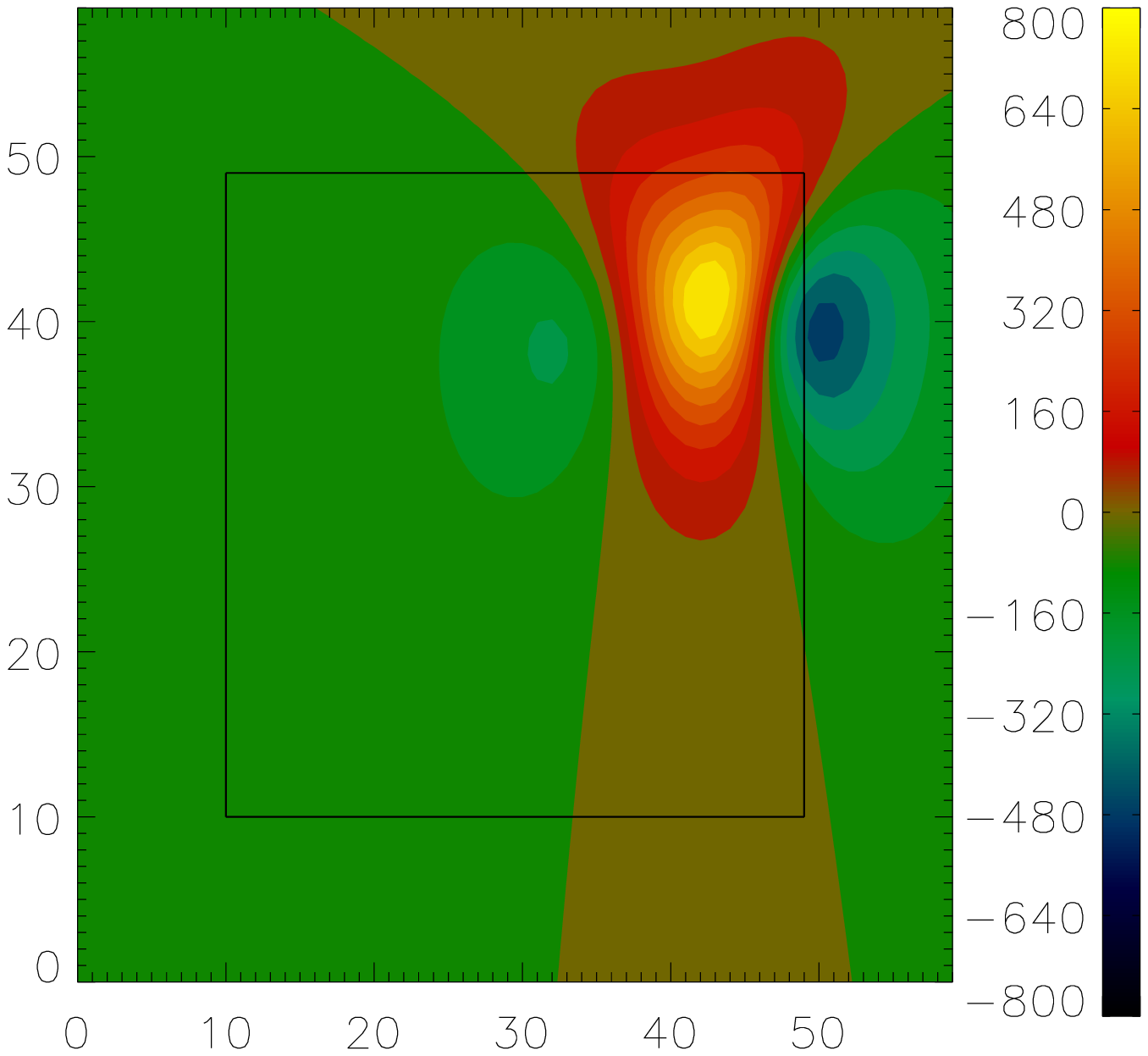}
}
\hspace*{\fill}
\caption{The pictures show artificial magnetograms extracted from the Low and Lou solution
with $l=0.5$ and $\phi=1.4$. The framed region corresponds to to the physical domain
with a resolution of $40 \times 40$ pixels. The computational box includes boundary layers
of $nd=10$ points towards each boundary. The colour-coding shows the normal magnetic field
strength on the photosphere. In the left hand picture (Force-Free I) the active region is centered and
in the right hand picture displaced (Force-Free II). The latter simulates data, where a significant
amount of flux is not balanced.}
\label{box1}
\end{figure}
\begin{figure}
\hspace*{\fill}
\mbox{\includegraphics[bb=45 20 445 260,clip,width=6.0cm]{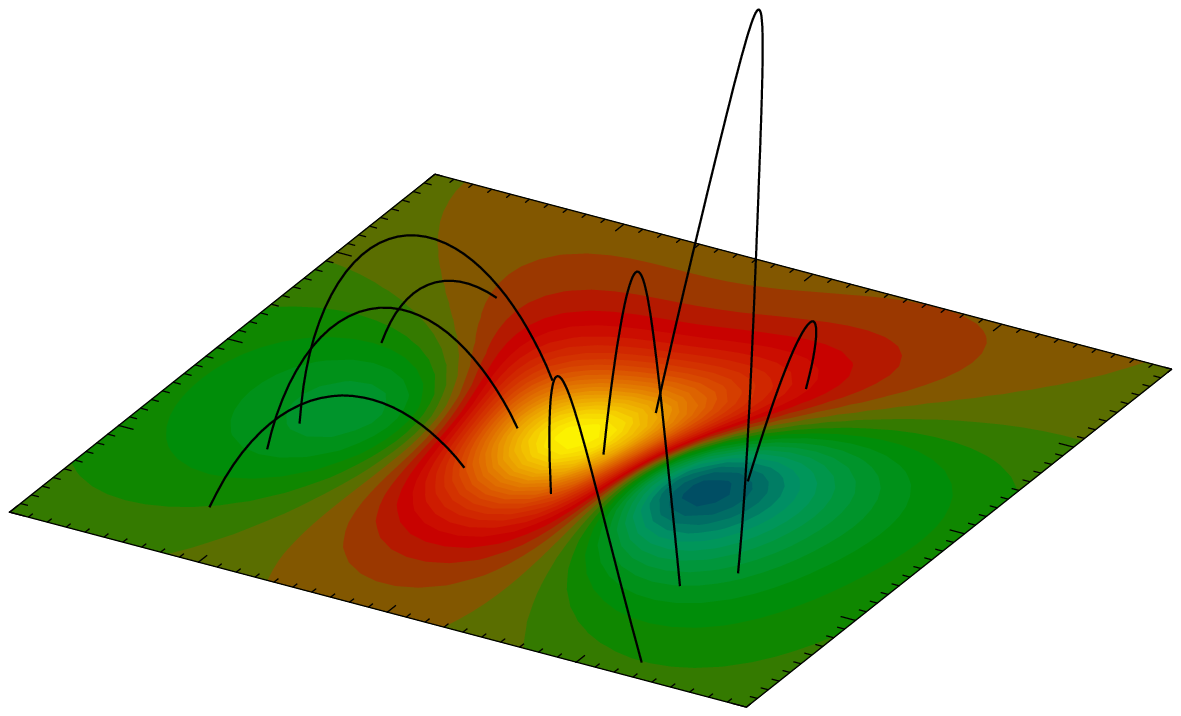}
\includegraphics[bb=45 20 445 260,clip,width=6.0cm]{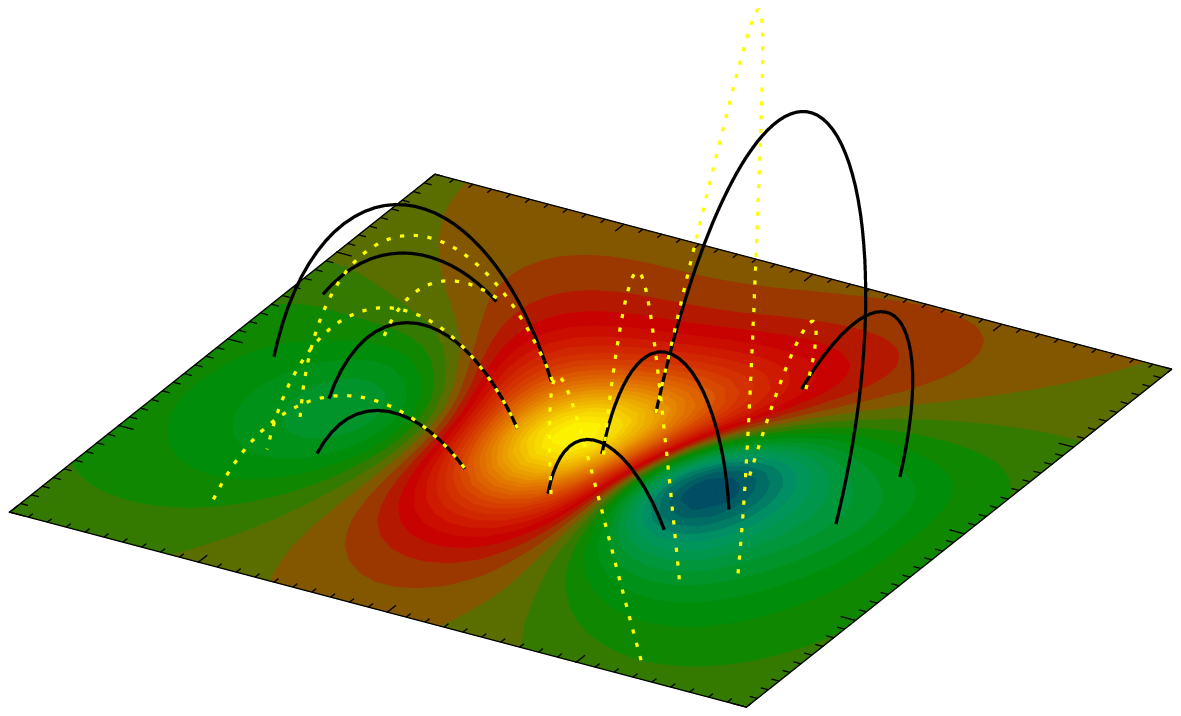}}
\mbox{\includegraphics[bb=45 20 445 260,clip,width=6.0cm]{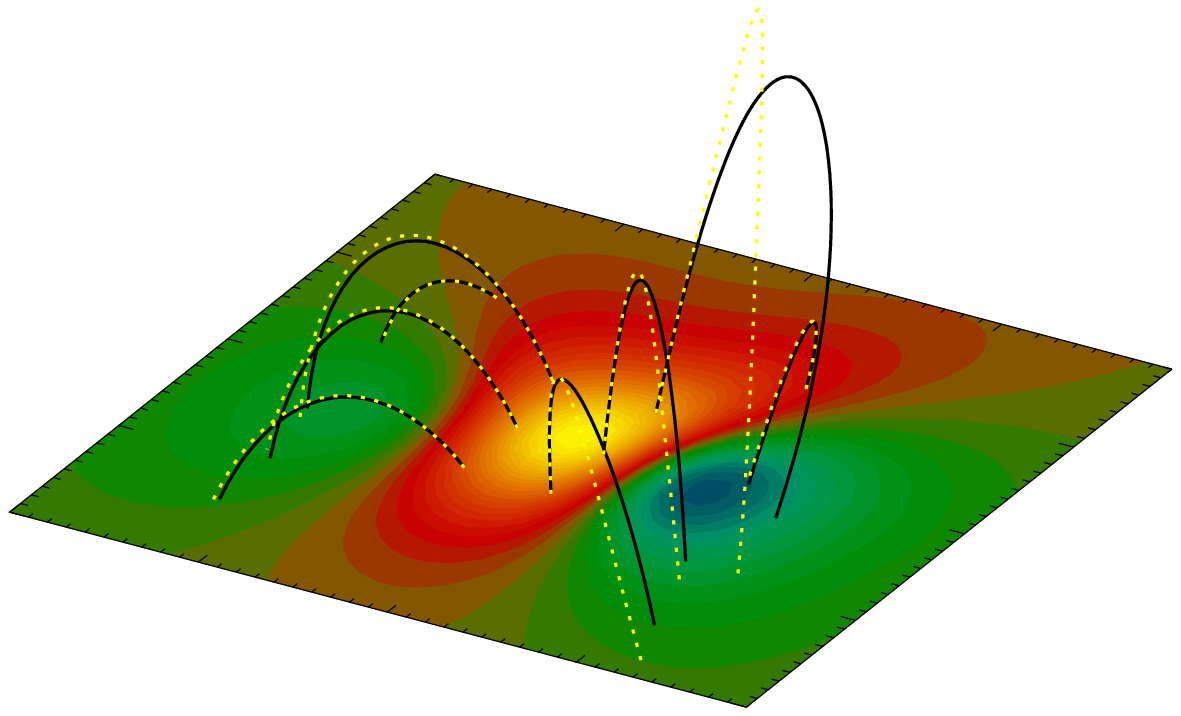}
\includegraphics[bb=45 20 445 260,clip,width=6.0cm]{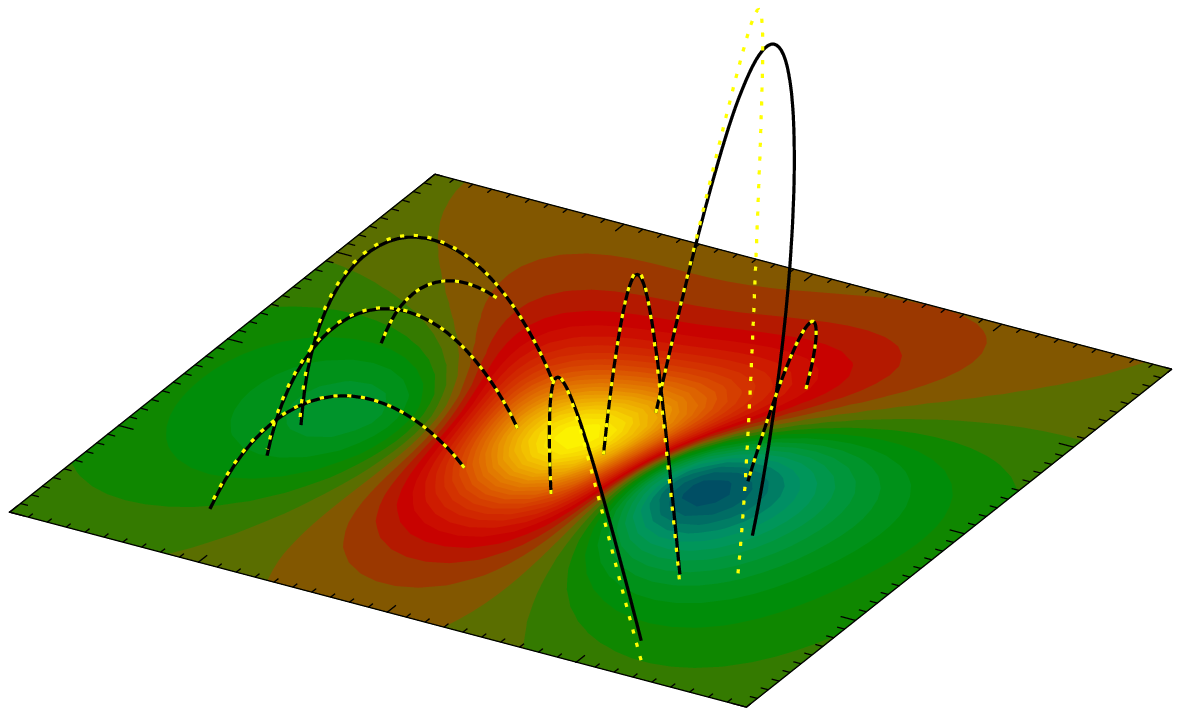}}
\mbox{\includegraphics[bb=45 20 445 260,clip,width=6.0cm]{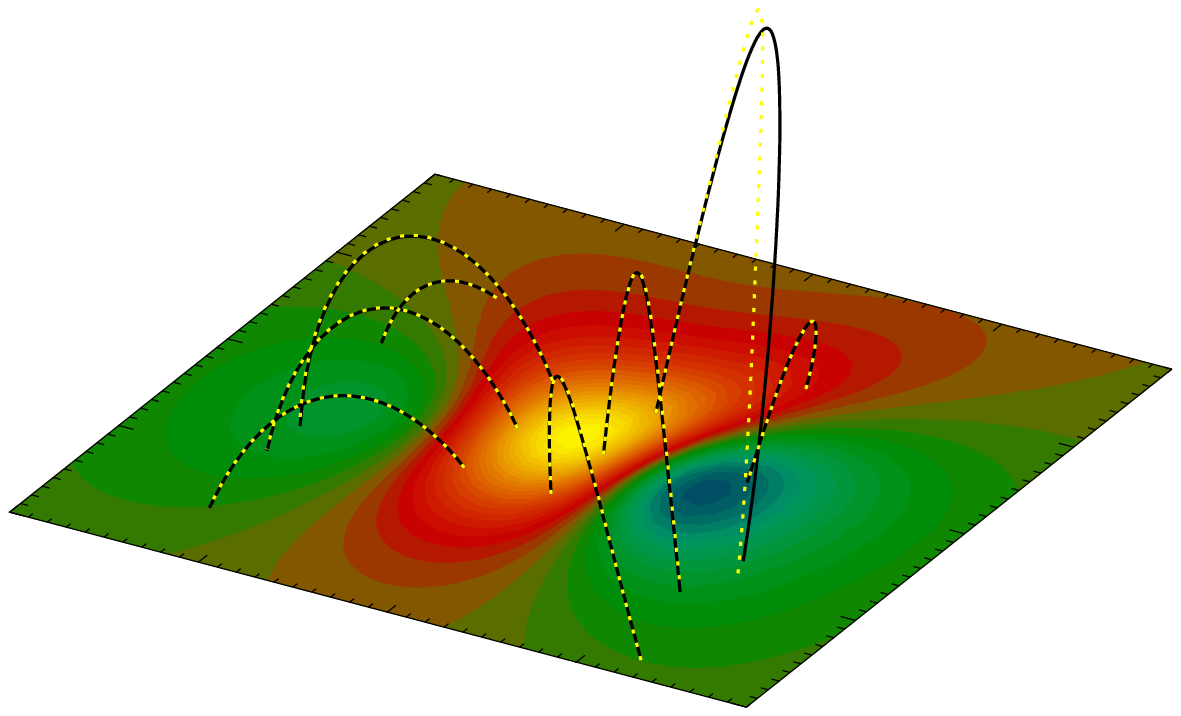}
\includegraphics[bb=45 20 445 260,clip,width=6.0cm]{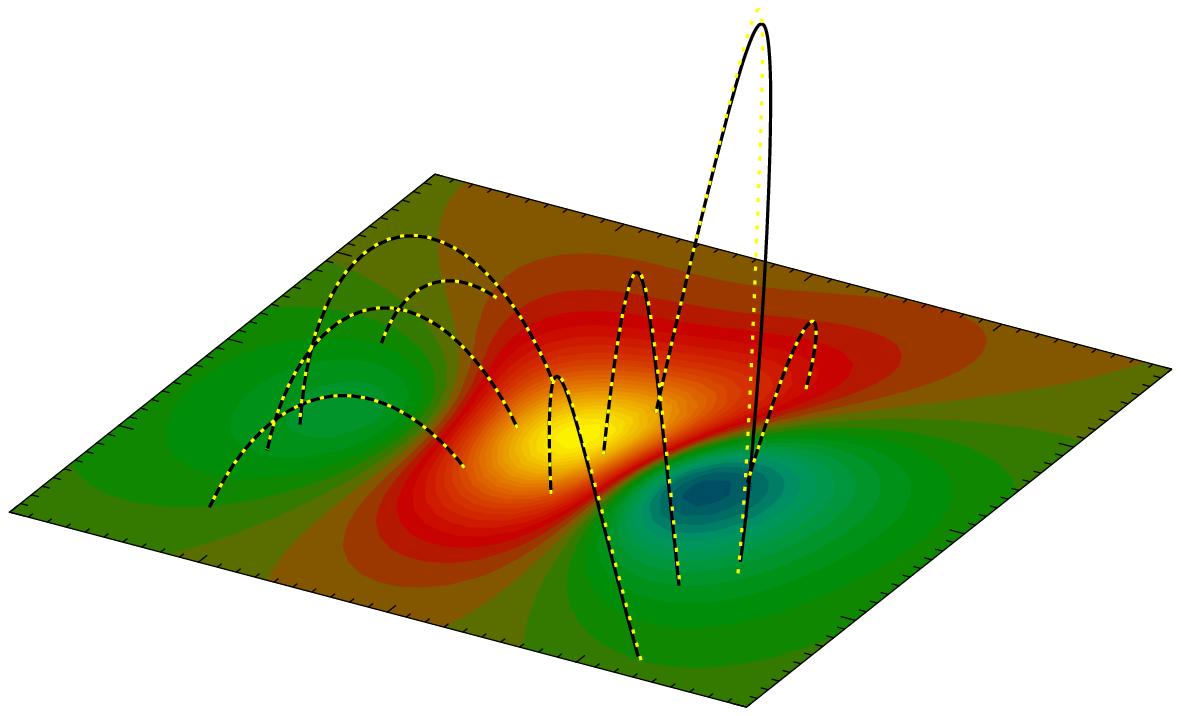}}
\hspace*{\fill}
\caption{Top row: The left hand side shows some field lines for the original Low and
Lou configuration ($l=0.5$ and $\phi=1.4$). The colour-coding shows the
normal magnetic field on the photosphere. We choose identical  inner footpoints
positions for all panels. The right hand side contains a
corresponding potential field reconstruction.  The yellow dotted field lines correspond
to the original Low and Lou solution in all pictures.
Center row: The left hand side shows a reconstruction without weighting function
and the right hand side with a linear weighting function and a 5 pixel boundary layer.
Bottom row: Both pictures show a reconstruction with cos-profile. In the left
hand picture a boundary layer of 10 points was used and in the right hand
picture a boundary layer of 20 points.}
\label{vergleich}
\end{figure}

To test our code we use a semi-analytic model active region
developed by \cite{lowandlou}.
We use the Low and Lou solution with $l=0.5$ and $\phi=1.4$ as test. The
normal photospheric magnetic field is normalized to a maximum of $800$ Gauss.
Figure \ref{box1} shows the normal magnetic field $B_z$ on the photosphere
for this configuration. The framed region contains the physical domain.
We investigate two cases, in Force-Free I the physical domain is
approximately flux-balanced and in Force-Free II not.
We are interested in reconstructing an inner region (physical domain) of
$40 \times 40 \times 20$ points and diagnose $L_i$ and  $|{\bf J} \times {\bf B}|$
averaged over the physical domain
here. We also diagnose $L$ in the whole computational box.

At the lateral and top boundary we introduce an
additional boundary layer of $nd$ points and $w$ decreases
from $1$ to $0$ in this layer. We investigate different profiles regarding
the weighting function, e.g.  linear, quadratic, cos and tanh
(see figure \ref{wprofil}). We
investigate how the size of the boundary layer influences the solution.
Table \ref{tab1} and figure \ref{vergleich} show the result of our
investigations.
%
\begin{table}
\caption{Details of runs to reconstruct force-free and non force-free magnetic fields.
All configurations have been calculated for a physical domain of $40 \times 40 \times 20$
grid points.
The first column specifies the profile of the boundary layer, the second
column the size $nd$ of the boundary layers, the third column the value of $L$ in
the computational box, the fourth column the value of $L_i$ in the physical domain and
the fifth column the force-free condition (for force-free configuration) or the
force balance (for MHS equilibria) averaged over the physical domain.
We specify the start-error and discretisation error for each configuration.}
\begin{tabular}{|r|r|r|r|r|}
\hline
Remarks & $nd$ & $\frac{ L}{ [{\rm T}^2 m]}$ & $\frac{\rm L_i}{ [{\rm T}^2 m]}$&
$\frac{|{\bf J} \times {\bf B}|}{[{\rm nN \, m}^{-3}]}$  \\
Force-Free I &&&& \\
Start-error &  &$9.5 \cdot 10^5$ &$ 9.5 \cdot 10^5$ & $5529$ \\
Discr-error &  &$3$ & $3$ &$12$  \\
& 0 &$2.6 \cdot 10^4$ & $2.6 \cdot 10^4$ &$391$  \\
b.r. & 0 &$2.3 \cdot 10^4$ & $2.3 \cdot 10^4$ &$383$  \\
lin & 3 &$2518 $ & $807 $ &$152 $ \\
lin & 5 &$1383 $ & $353 $ &$112 $ \\
lin &10 &$524 $ & $92 $ &$65 $  \\
quad &10 &$800 $ & $98 $ &$67 $  \\
cos &10 &$334 $ & $65 $ &$55 $ \\
tanh &10 &$468 $ & $73 $ &$59 $  \\
hom &10 &$8052 $ & $193 $ &$90 $  \\
lin &20  &$329 $ & $36 $ &$45 $  \\
quad &20  &$588 $ & $64 $ &$62 $ \\
cos &20  &$240 $ & $27 $ &$37 $ \\
tanh &20  &$291 $ & $27 $ &$37 $ \\
\hline
Force-free II &&&& \\
Start-error&  & $5.9 \cdot 10^5$ &$5.9 \cdot 10^5$ & $3852$ \\
Discr-error &  &$9$ & $9$ &$11$ \\
& 0 &$1.3 \cdot 10^5$ & $1.3 \cdot 10^5$ &$1280$   \\
b.r. & 0 &$7.2 \cdot 10^4$ & $7.2 \cdot 10^4$ &$1082$   \\
cos & 10 &$759$ & $166$ &$85$ \\
hom & 10 &$2.4 \cdot 10^4$ & $285$ &$105$ \\
cos & 20 &$624$ & $62$ &$52$  \\
\hline
MHS &&&&$\frac{ \rm Force-balance}{ [{\rm nN \, m}^{-3}]}$  \\
Start-error &  & $2.7 \cdot 10^5$ &$2.7 \cdot 10^5$ & $2935$ \\
Discr-error & &$45$ & $45$ &$34$  \\
    &0 & $4.2 \cdot 10^4$ &$4.2 \cdot 10^4$ &$1022$ \\
cos & 10 & 2000 & 401 & $99$ \\
cos & 20 & 1161 & 153 & $46$ \\
\hline
\end{tabular}
\label{tab1}
\end{table}
\subsection{Force-Free I}
Figure \ref{vergleich} top right panel shows that the field lines of
a potential field reconstruction are clearly different from the original Low
and Lou solution. This naturally leads to high values of the functional $L_i$
and large ${\bf J} \times {\bf B}$ forces after the bottom boundary has been
replaced by the original vector magnetogram. We first apply the optimization
code without weighting function $(nd=0)$. Here the boundaries of the physical domain
coincide with the computational boundaries. The lateral and top boundary
have the value of the potential field during the iteration. Some low lying field lines are
represented quite well (left hand picture in figure \ref{vergleich} second row).
These field lines close to the box center are
of course close to the bottom boundary and far away from the other
boundaries. The (observed) bottom boundary has a higher influence on the field here
than the potential lateral and top boundary.
Other field lines, especially high reaching field lines deviate from the analytic solution
(yellow dotted line).

The values $L_i$ and
$|{\bf J} \times {\bf B}|$  provide a quantitative measure of the
quality of the reconstructed magnetic field in the physical domain. High values correspond to
a significant deviation from the force-free state. We applied the method of
boundary relaxation (marked with {\it b.r.} in table \ref{tab1}), but the result
only slightly improved  \footnote{The boundary relaxation method uses
the iterative improvement $\frac{\partial {\bf B}}{\partial t}= \tilde{\mu} {\bf
G}$ for fields on the lateral and top boundary with ${\bf G}$ as in (\ref{defG}) in
addition to (\ref{iterateB}) with $w=1$.
See (\cite{twtn03} section 4.2.3 for details.}.

We investigate how the size and
shape of a boundary layer influences the quality of the reconstruction. Both
the comparison of the field lines (figure \ref{vergleich}
panel 4 to 6)  as well as the quantitative values in table \ref{tab1}
show that the quality of the reconstruction improves significantly
with the size of the boundary layer (thickness in number of grid points $nd$).
The larger
computational box displaces the lateral and top boundary further away from
the physical domain and consequently its influence on the solution decreases.
As a result the magnetic field in the physical domain is
dominated by the vector magnetogram data, which is exactly what is required.
We find that a cos-profile of the weighting function provides the best
results, followed closely by a tanh-profile. The main advantage of these
profiles seem to be that they have smooth gradients at the boundary of the
physical domain to the boundary layer as well as at the boundaries of the
computational box. We tried also a homogenous profile ($w=1$ in the whole
computational box, marked with {\it hom} in table \ref{tab1}). In the homogeneous
case $w$ is equal $1$ in the whole computational box and a  boundary layer does not exist.
The use of $nd=10$ for the homogeneous case in table \ref{tab1} is only for
diagnostic reasons and indicates that $Li$ and $|{\bf J} \times {\bf B}|$
have been computed in the same interior box (physical domain) as in the cases
with weighting function for comparison.
The effect that the lateral and top boundary are far away
from the physical domain remains valid here, but the use of
a weighting function in the boundary layers provides much better results.
\subsubsection{The influence of noise}
The previous calculations have been carried out under the assumption that the magnetic
field on the boundary of the computational box is known exactly. Such an idealized situation
will not be found when real vector magnetogram data are used.
To keep control over the amount of uncertainty, we have carried
out test runs by adding random noise to the vector magnetogram.
We add the noise by multiplying the exact
boundary conditions with a number $1+\delta$ where $\delta$ is
a random number in the range $-n_l \leq \delta \leq n_l$
and $n_l$ is the noise level. We investigate the effect of noise for
Force-Free I with a boundary layer of $nd=10$ grid points and a cos
profile in $w$ for different noise levels.
\begin{figure}
\hspace*{\fill}
\mbox{\includegraphics[bb=45 20 445 260,clip,width=6.0cm]{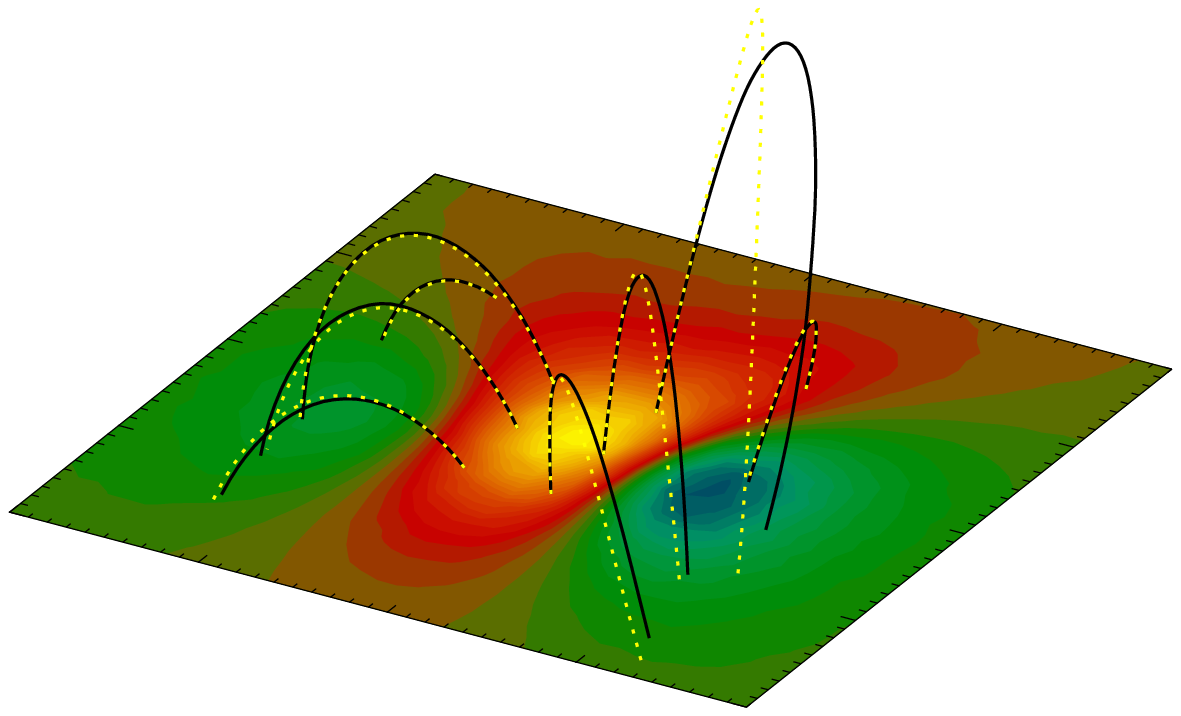}
\includegraphics[bb=45 20 445 260,clip,width=6.0cm]{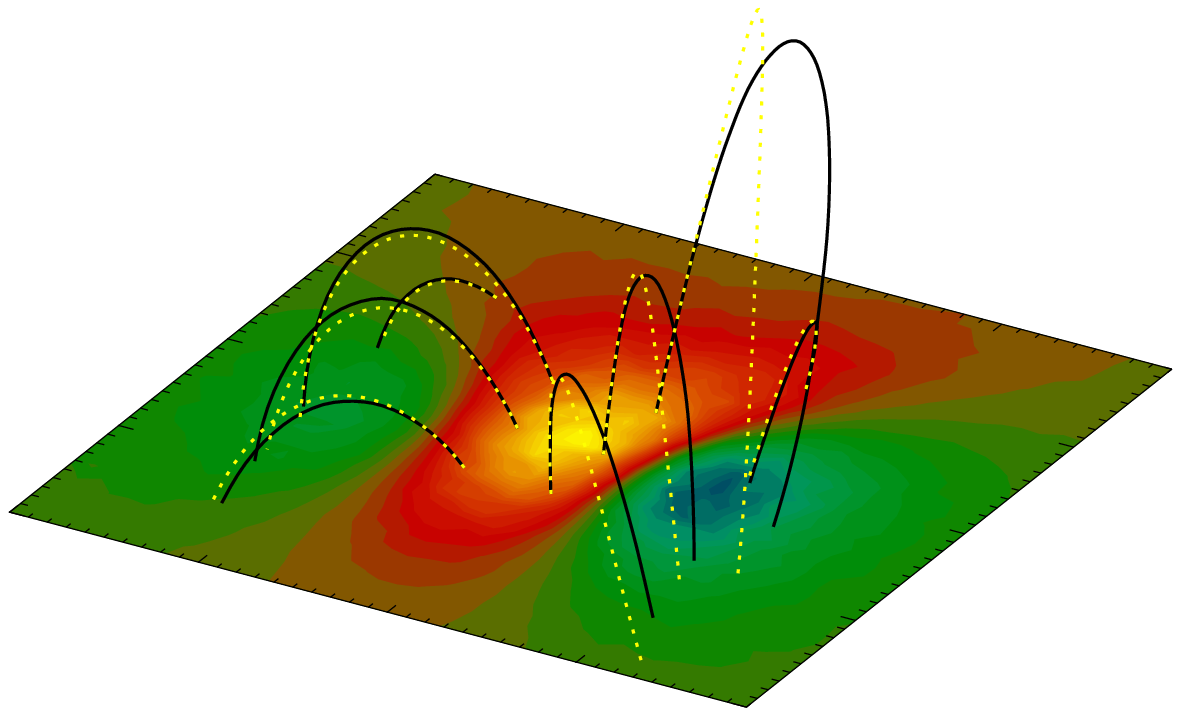}}
\mbox{\includegraphics[clip,height=5.0cm]{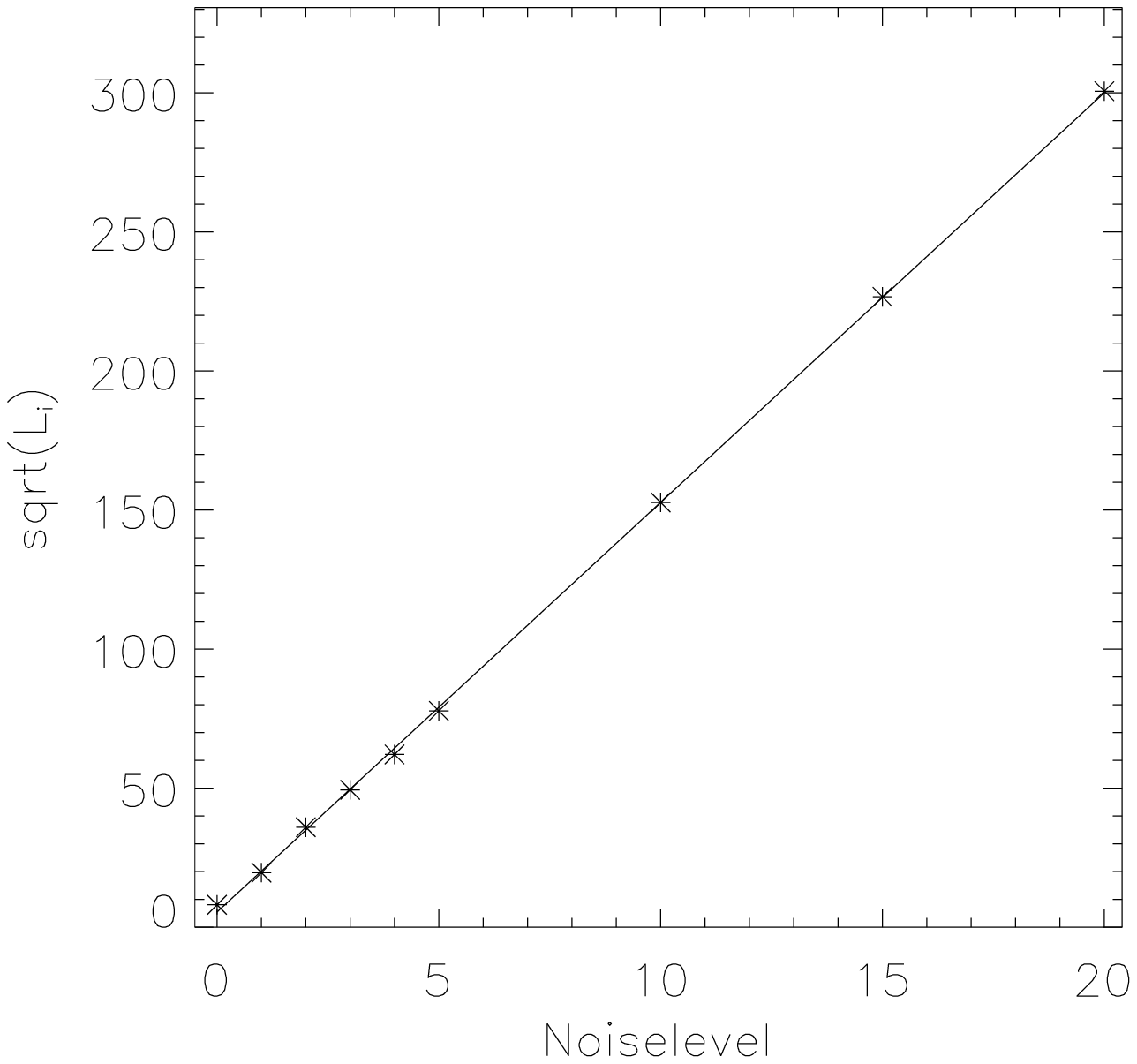}
\includegraphics[clip,height=5.0cm]{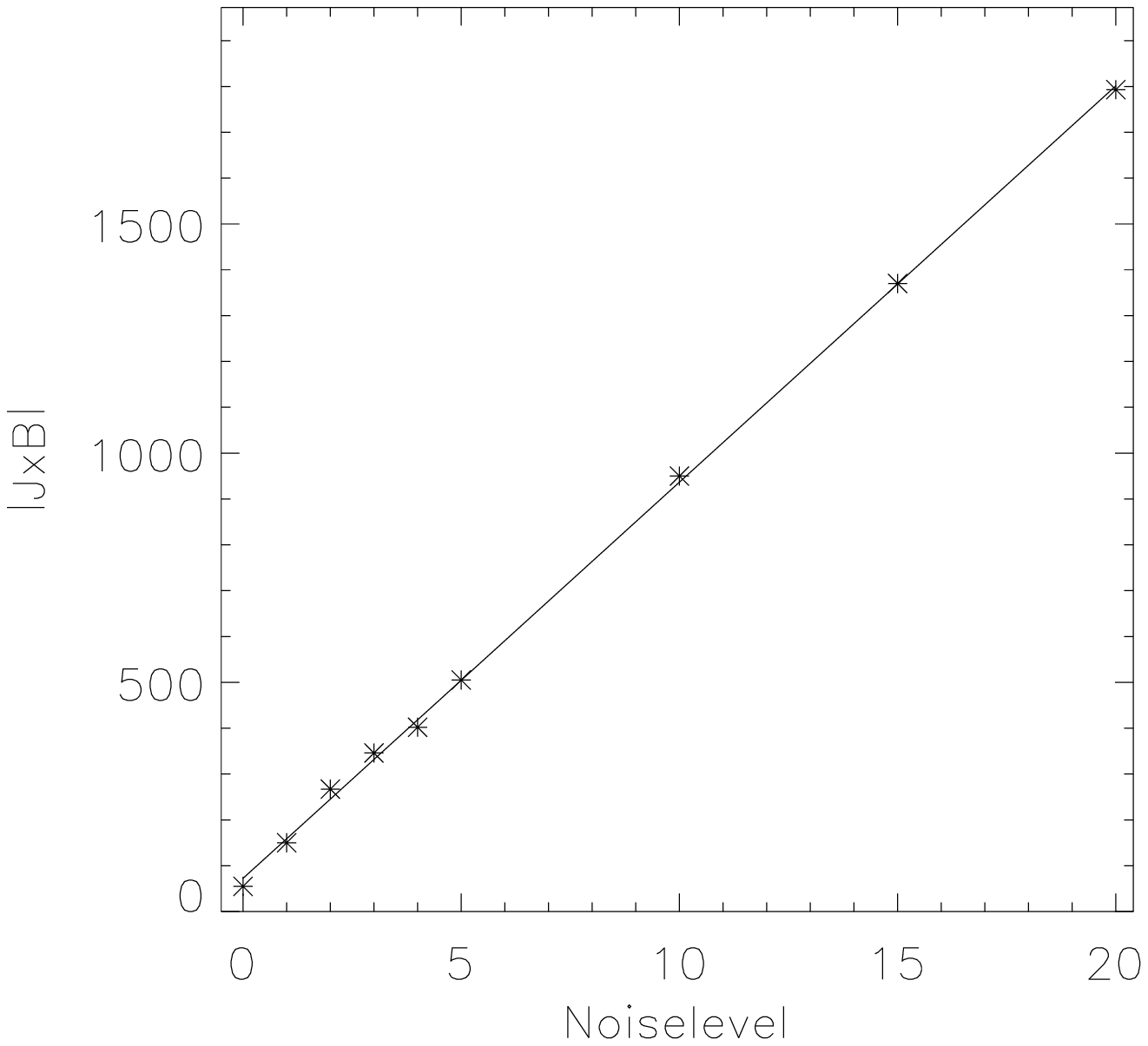}}
\hspace*{\fill}
\caption{Top row: The left hand side shows some field lines (same footpoints as in figure
\ref{vergleich}) for a noise level of $5 \%$ and the right hand side for a noise
level of $10 \%$. The yellow dotted field lines correspond
to the original Low and Lou solution. We used a
boundary layer of $nd=10$ grid points and a cos profile in $w$.
Bottom row: The left hand side shows a plot of the noise level against $\sqrt{L_i}$ and the
right hand side a plot of the noise level against $|{\bf J} \times {\bf B}|$. The stars correspond
to the data points and the line is a linear fit.}
\label{noisefigure}
\end{figure}

\begin{table}
\caption{The influence of noise.
All configurations have been calculated for a physical domain of $40 \times 40 \times 20$
grid points, a boundary layer of $nd=10$ grid points and a cos profile in $w$.
The first column specifies the noise level, the second
column the value of $L$ in
the computational box, the third column the value of $L_i$ in the physical domain and
the forth column the force-free condition averaged over the physical domain.}
\begin{tabular}{|r|r|r|r|}
\hline
noise level & $\frac{ L}{ [{\rm T}^2 m]}$ & $\frac{\rm L_i}{ [{\rm T}^2 m]}$&
$\frac{|{\bf J} \times {\bf B}|}{[{\rm nN \, m}^{-3}]}$  \\
no noise &$334 $ & $65 $ &$55 $ \\
$ 1 \%$   &$678 $ & $385$ &$150$ \\
$ 2 \%$   &$1674 $ & $1294 $ &$267 $ \\
$ 3 \%$   &$2847 $ & $2440 $ &$346 $ \\
$ 4 \%$   &$4243 $ & $3858 $ &$402 $ \\
$ 5 \%$   &$6521 $ & $6048 $ &$505 $ \\
$10 \%$   &$24240 $ & $23332 $ &$950 $ \\
$15 \%$   &$52827 $ & $51382 $ &$1370 $ \\
$20 \%$   &$92535 $ & $90360 $ &$1793 $ \\
\hline
\end{tabular}
\label{tab2}
\end{table}
Table \ref{tab2} and figure \ref{noisefigure} show our results.
$\sqrt{L}$, $\sqrt{L_i}$ and $|{\bf J} \times {\bf B}|$ increase linearly
with the noise level. The field line pictures in the upper
panel of figure \ref{noisefigure} show that low lying field lines are
represented almost correctly while there are some deviations from the
analytic Low and Lou solution for high reaching field lines. As the noise is
completely random and independent between neighbouring grid points there are
obvious difficulties regrading the discretisation. Our code uses forth order
finite differences and consequently five grid points are required to compute
derivatives. The highly oscillatory  noise (The spatial variation of the
noise corresponds to the spatial resolution of the grid.) naturally results in finite
gradients which are linearly dependent from the noise level. $|{\bf J} \times {\bf B}|$
is linearly dependent from  spatial deviations and does consequently also
depend linearly on the noise level. The computation of $L$ and $L_i$ contain gradients
squared and are consequently quadratic dependent from the noise level. A
pre-processing of raw magnetogram data, e.g. a Fourier filter or some
smoothing might help to reduce the effect of random noise.

As the weighting function is designed to diminish the effect of the lateral
and top boundary towards the solution, we investigate here how changes on these
boundaries influence the solution. We undertook (for $nd=10$ and a cos-profile)
two runs with  linear force free magnetic fields ($\alpha L_x=2.0$ and
$\alpha L_x=-2.0$) on the lateral and top boundaries. $L$,$L_i$ and
$|{\bf J} \times {\bf B}|$ are of the
same order as for potential boundary conditions:
\begin{tabular}{llll}
potential  : & $L=334$,  & $L_i=65$,& $|{\bf J} \times {\bf B}|=55$ \\
$\alpha L_x=2$ :& $L=458$, & $L_i=92$,& $|{\bf J} \times {\bf B}|=75$ \\
$\alpha L_x=-2$ :& $L=492$, & $L_i=62$,& $|{\bf J} \times {\bf B}|=59$ \\
\end{tabular}

The magnetic field lines calculated from these fields look exactly the
same as for potential boundary conditions. We conclude that
the influence of the lateral and top boundary conditions towards the solution
in the physical domain is indeed very small.
\subsection{Force-Free II}
The magnetogram on the right hand side of figure \ref{box1} is obviously bad conditioned.
Significant parts of the magnetic flux are outside of the framed physical
domain. In principle it would be better to always choose a magnetogram
such that the majority of the flux is centered and the overall flux is
approximately balanced but due to the limited size of
vector magnetograms  sometimes vector magnetograph data are not available
over an entire active region. Here we investigate how this influences the
quality of the reconstruction. We find that the reconstruction without
weighting function provides much worse results in such a case than for
a well centered
force-free configuration. This is obvious caused by the potential field
assumed on the lateral
boundary  in a region of a high magnetic flux and current density.
The influence of the inconsistent lateral boundary conditions on the magnetic
field in the physical domain is too strong. Here the method of boundary
relaxation improves the quality of $L_i$ nearly by a factor of two, but the
result still remains unsatisfactory. For runs with weighting function the result
is basically similar as for {Force-Free I}. A large boundary layer (now
including all active parts of the magnetogram) improves the quality of the
reconstruction significantly. For the largest boundary layer $nd=20$ points wide
the quality of the reconstruction is approximately equal to a $10$ point wide
boundary layer in the Force-Free I case.

If for observational data only parts
of an active region are available as vector magnetogram data a nonlinear
reconstruction of the coronal magnetic field might become difficult or
impossible. One could try to get the corresponding normal component of the
photospheric magnetic field from other sources, e.g. the line of sight magnetograph
MDI on SOHO and make assumptions regarding the transversal magnetic field.
The reconstructed magnetic field will then of course be influenced by these
(not observed) assumptions.
\section{Tests for non force-free configurations}
\label{test2}
\begin{figure}
\hspace*{\fill}
\mbox{\includegraphics[bb=45 20 445 180,clip,width=6.0cm]{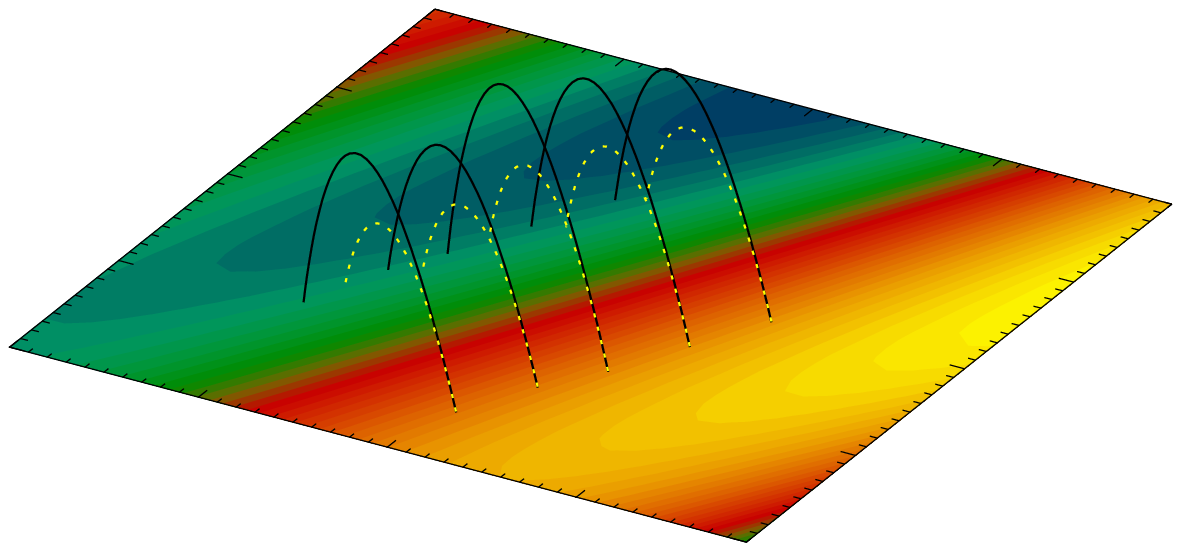}
\includegraphics[bb=45 20 445 180,clip,width=6.0cm]{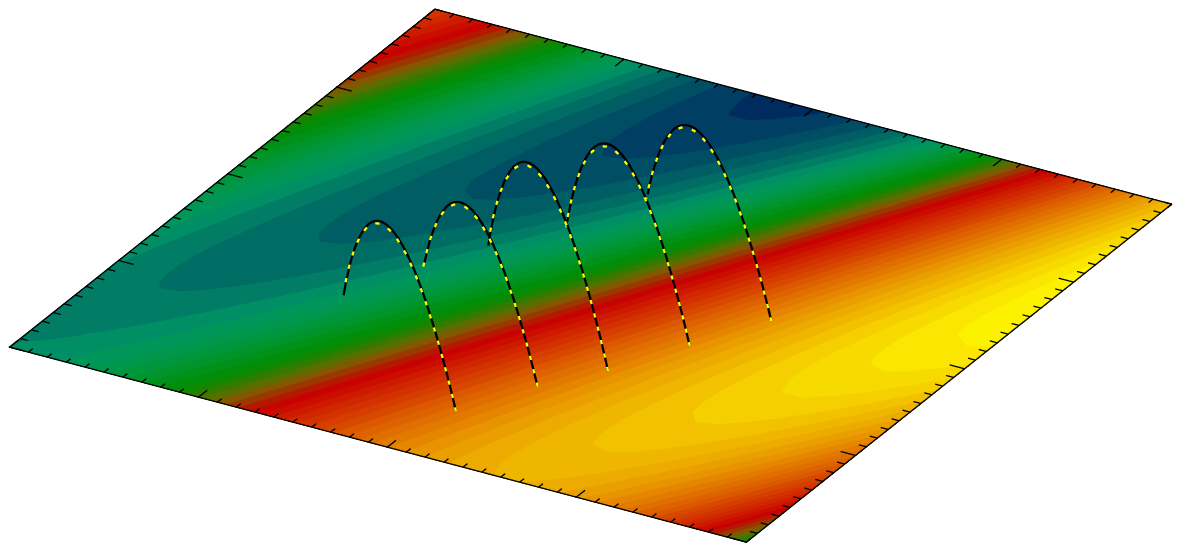}}
\caption{Here we illustrate an example of a non force-free reconstruction.
The left hand side shows a potential field reconstruction and
the right hand side a MHS reconstruction with help of our optimization-code
and a boundary layer of 10 points with cos-profile. The yellow dotted lines show
corresponding field lines for the exact solution. The colour-coding shows the normal
magnetic field strength on the photosphere. }
\label{vergleich2}
\end{figure}
For finite $\beta$ configurations it is necessary to include
pressure and gravity forces and solve
the magneto hydro static equations (\ref{forcebal})-(\ref{solenoidal}).
To test our code we use an analytic MHS equilibrium (see \cite{twbi03} section 3.1.)
with an average plasma $\beta=0.2$. From this analytic solution we extract a
photospheric vector magnetogram and the coronal density distribution. It is
convenient to use the analytic density distribution here to test
our code, but one has to keep in mind that for observational data the coronal density
structure has to be reconstructed with help of a tomographic inversion.
Corresponding observational data (line of sight integrals of the coronal density
structure) are expected from the Stereo-mission. The tomographic inversion of
these data is a challenging problem on its own and we are not going to discuss it
here. \cite{twbi03} give an overview over the tomographic inversion
algorithm. Here we consider the vector magnetogram and the plasma density
distribution as given.
The left hand panel of figure \ref{vergleich2} shows a potential magnetic
field reconstruction consistent with the normal component of the magnetic
field extracted from an analytic magneto hydro static equilibrium. The
comparison of the magnetic field lines of the potential field with the exact
field show significant deviations. In table \ref{tab1} we diagnose $L$ and
$L_i$ similarly as for the force-free cases and we diagnose the force balance
$|{\bf J} \times {\bf B}-\nabla P|$  here.  Similar as
in the force-free case we find that the quality of the reconstruction is poor
for an optimization without weighting function. A boundary layer of $nd=10$
points together with a cos-profile of $w$ improves the value of functional $L_i$ by a
factor of $10^{-2}$ in the physical domain and the force balance
\footnote{The functional $L_i$ contains the square of the force balance, see
the definition \ref{defL2}.} by a factor of
about $10^{-1}$. In the right hand panel of figure \ref{vergleich2}
we compare the reconstructed field with the analytic solution and both
coincide within the graphical resolution. A larger boundary layer of $nd=20$
further improves the
quantitative measures of the non force-free reconstruction.
The influence of a weighting function acts similarly for force-free and non
force-free configuration. In both cases the reconstruction result is
significantly improved.
\section{Conclusions}
\label{conclusions}
In this paper we improved the optimization method for the reconstruction of
non linear force-free and not force-free coronal magnetic fields.
The optimization method minimizes a functional which consists of a
quadratic form of the force balance and the solenoidal condition.
While in previous optimization attempts the magnetic field needed to be described on
all six boundaries of a computational box, our approach allows us to reconstruct
the coronal magnetic field from the bottom boundary data alone. This is possible by
the introduction of a boundary layer around the physical domain. The physical
domain is a cubic area within which we want to reconstruct the coronal
magnetic field consistent with photospheric vector magnetogram data.
 The boundary layer replaces the hard lateral and top boundary used previously.
We showed that the limit of an infinitesimally thin boundary layer
formally coincides with the original hard boundary. However, our test
calculations show that a finite size weighted boundary yield much better results.

We introduced a weighting function. In the physical domain
the weighting function is unity. It drops monotonically  in the boundary
layer and reaches zero at the boundary of the computational box. At the boundary
of the computational box we set the field to the value of the
potential field computed from  $B_n$ at the bottom boundary.
This choice is not only convenient
due to its  mathematic simplicity, but as well physical
motivated. The coronal magnetic field is approaching a potential field in quite
regions and  high in the corona.
The current program uses a cartesian geometry and is designed to reconstruct
basically isolated active regions. The photospheric magnetic field outside this
active region (and a surrounding area) is ignored. This approach is the better justified
the more the active region is isolated. Often, however, active regions are not always completely
isolated but magnetically connected with other active regions. To include this effect
it would in principle be preferable  to reconstruct the
complete coronal magnetic field with vector magnetogram data on the whole photosphere as
boundary. This would avoid the problems of prescribing the lateral boundaries
and only the top boundary has to be chosen accordingly, similar to the source
surface of potential field reconstructions. We do not expect any difficulties
to apply our method to problems in spherical geometry.
Unfortunately the required photospheric boundary data are not available because current
vector magnetographs do not observe the entire solar surface. The situation
will probably significantly improved by the vector spectromagnetograph of the
the SOLIS project. It will deliver full disk vector magnetograms.
\acknowledgements
We thank Bernd Inhester for useful discussions.
This work was supported by  DLR-grant 50 OC 0007.
We thank the referee Thomas R. Metcalf for useful remarks.
\appendix
\renewcommand{\theequation}{\Alph{section}.\arabic{equation}}
\section{Derivation of ${\bf \tilde{F}}$ and ${\bf \tilde{G}}$ in (\ref{minimize1}).}
\label{appendixA}
\setcounter{equation}{0}

\begin{equation}
L=\int_{V} \; w(x,y,z) \; B^2 \; (\Omega_a^2+\Omega_b^2) \; d^3x
\end{equation}
\begin{eqnarray}
{\bf \Omega_a} &=& B^{-2} \;\left[(\nabla \times {\bf B}) \times {\bf B} + {\bf u} \right]
\nonumber\\
{\bf \Omega_b} &=& B^{-2} \;\left[(\nabla \cdot {\bf B}) \; {\bf B} \right]
\nonumber\\
{\bf u} &=&-\mu_0 (\nabla P +\rho \nabla \Psi)
\end{eqnarray}
We vary L with respect to an iteration parameter t and get
\begin{eqnarray}
\frac{1}{2} \; \frac{d L}{d t} &=&
\int_{V} \; w \; {\bf \Omega_a} \cdot \frac{\partial}{\partial t}
[(\nabla \times {\bf B}) \times {\bf B} + {\bf u} ]  \; d^3x
\nonumber\\
&+&\int_{V} \; w \; {\bf \Omega_b} \cdot  \frac{\partial}{\partial t}
[(\nabla \cdot {\bf B}) \; {\bf B}] \; d^3x
\nonumber\\
&-& \int_{V} \; w \; (\Omega_a^2 + \Omega_b^2) \;
{\bf B} \cdot \frac{\partial {\bf B}}{\partial t} \; d^3x
\end{eqnarray}
Our aim is now to use vector identities and Gauss law in such way that all
terms contain a product with $\frac{\partial {\bf B}}{\partial t}$. This will
allow us to provide explicit evolution equations for ${\bf B}$ to minimize $L$.
The third term has the correct form already. We expand the first and second term
\begin{eqnarray}
\Rightarrow \frac{1}{2} \; \frac{d L}{d t} &=&
\int_{V}\; w \; {\bf \Omega_a} \cdot \left[ \left(\nabla \times \frac{\partial {\bf B}}{\partial t}\right) \times {\bf B} \right]  \; d^3x
\nonumber\\
&+& \int_{V} \; w \; {\bf \Omega_a} \cdot
\left[(\nabla \times {\bf B}) \times \frac{\partial {\bf B}}{\partial t}\right]   \; d^3x
\nonumber\\
&+&\int_{V} \; w \; {\bf \Omega_b} \cdot
\left[ \left(\nabla \cdot \frac{\partial {\bf B}}{\partial t} \right) \; {\bf B} \right] \; d^3x
\nonumber\\
&+&\int_{V} \; w \; {\bf \Omega_b} \cdot
\left[(\nabla \cdot {\bf B}) \; \frac{\partial {\bf B}}{\partial t} \right] \; d^3x
\nonumber\\
&-& \int_{V} \; w \; (\Omega_a^2 + \Omega_b^2) \; {\bf B} \cdot \frac{\partial {\bf B}}{\partial t} \; d^3x
\end{eqnarray}
The fourth and fifth term have the correct form. We apply the vector
identities
${\bf a} \cdot ({\bf b} \times {\bf c})={\bf b} \cdot ({\bf c} \times {\bf a})=
{\bf c} \cdot ({\bf a} \times {\bf b})$
to the first and second term
\begin{eqnarray}
\Rightarrow \frac{1}{2} \; \frac{d L}{d t} &=&
\int_{V} \; w \; \left(\nabla \times \frac{\partial {\bf B}}{\partial t}\right) \cdot ({\bf B} \times  {\bf \Omega_a})  \; d^3x \nonumber\\
&+& \int_{V} \; w \; \frac{\partial {\bf B}}{\partial t} \cdot
({\bf \Omega_a} \times (\nabla \times {\bf B}) \;     \; d^3x \nonumber\\
&+&\int_{V} \; w \; ({\bf \Omega_b} \cdot {\bf B}) \; \nabla \cdot  \frac{\partial {\bf B}}{\partial t}
\; d^3x \nonumber\\
&+&\int_{V} \; w \; \left[{\bf \Omega_b} \;
(\nabla \cdot {\bf B})\right] \cdot \frac{\partial {\bf B}}{\partial t}  \; d^3x \nonumber\\
&-& \int_{V} \; w \; (\Omega_a^2 + \Omega_b^2) \; {\bf B} \cdot \frac{\partial {\bf B}}{\partial t} \;
d^3x
\end{eqnarray}
Term two, four and five have the correct form. We apply
$(\nabla \times {\bf a}) \cdot {\bf b}={\bf a} \cdot (\nabla \times {\bf
b})+\nabla \cdot ({\bf a} \times {\bf b})$ to term 1 and
$\psi \nabla \cdot {\bf a}=-{\bf a} \cdot \nabla \psi+\nabla \cdot({\bf a \psi})$
to term 3
\begin{eqnarray}
\Rightarrow \frac{1}{2} \; \frac{d L}{d t} &=&
-\int_{V}\; w \; \frac{\partial {\bf B}}{\partial t} \cdot [\nabla \times ({\bf \Omega_a} \times {\bf B} )]  \; d^3x \nonumber\\
&-& \int_{V}\; w \; \nabla \cdot \left[({\bf \Omega_a} \times {\bf B} ) \times \frac{\partial {\bf B}}{\partial t}  \right]  \; d^3x \nonumber\\
&+& \int_{V}\; w \; \frac{\partial {\bf B}}{\partial t} \cdot
({\bf \Omega_a} \times (\nabla \times {\bf B}) \;     \; d^3x \nonumber\\
&-&\int_{V}\; w \; \nabla ({\bf \Omega_b} \cdot {\bf B}) \;  \cdot  \frac{\partial {\bf B}}{\partial t}
\; d^3x \nonumber\\
&+&\int_{V}\; w \; \nabla \cdot \left[ ({\bf \Omega_b} \cdot {\bf B}) \; \frac{\partial {\bf B}}{\partial
t} \right]
\; d^3x \nonumber\\
&+&\int_{V}\; w \; \left[{\bf \Omega_b} \;
(\nabla \cdot {\bf B})\right] \cdot \frac{\partial {\bf B}}{\partial t}  \; d^3x \nonumber\\
&-& \int_{V}\; w \;  (\Omega_a^2 + \Omega_b^2) \; {\bf B} \cdot \frac{\partial {\bf B}}{\partial t} \; d^3x
\end{eqnarray}
Until here the derivation has been identical with the method without weighting function
\cite{twbi03}. Now we get additional terms with respect to the weighting function.
We apply $\psi \nabla \cdot {\bf a}=\nabla \cdot(\psi {\bf a})-{\bf a} \cdot \nabla \psi$ to
term 2 and 5
\begin{eqnarray}
\Rightarrow \frac{1}{2} \; \frac{d L}{d t} &=&
-\int_{V}\; w \; \frac{\partial {\bf B}}{\partial t} \cdot [\nabla \times ({\bf \Omega_a} \times {\bf B} )]  \; d^3x \nonumber\\
&-& \int_{V} \nabla \cdot \left[\; w \; ({\bf \Omega_a} \times {\bf B} ) \times \frac{\partial {\bf B}}{\partial t}
\right]  \; d^3x \nonumber\\
&+& \int_{V}  \left[({\bf \Omega_a} \times {\bf B} ) \times \frac{\partial {\bf B}}{\partial t}
\right] \cdot \nabla w  \; d^3x \nonumber\\
&+& \int_{V}\; w \; \frac{\partial {\bf B}}{\partial t} \cdot
({\bf \Omega_a} \times (\nabla \times {\bf B})) \;     \; d^3x \nonumber\\
&-&\int_{V}\; w \; \nabla ({\bf \Omega_b} \cdot {\bf B}) \;  \cdot  \frac{\partial {\bf B}}{\partial t}
\; d^3x \nonumber\\
&+&\int_{V} \nabla \cdot \left[\; w \; ({\bf \Omega_b} \cdot {\bf B}) \; \frac{\partial {\bf B}}{\partial
t} \right] \; d^3x \nonumber\\
&-&\int_{V} \left[ ({\bf \Omega_b} \cdot {\bf B}) \; \frac{\partial {\bf B}}{\partial
t} \right]\cdot \nabla w  \; d^3x \nonumber\\
&+&\int_{V}\; w \; \left[{\bf \Omega_b} \;
(\nabla \cdot {\bf B})\right] \cdot \frac{\partial {\bf B}}{\partial t}  \; d^3x \nonumber\\
&-& \int_{V}\; w \;  (\Omega_a^2 + \Omega_b^2) \; {\bf B} \cdot \frac{\partial {\bf B}}{\partial t} \; d^3x
\end{eqnarray}
The terms 1,4,5,7,8 and 9 have the correct form. We apply Gauss' law to term 2 and 6
\begin{eqnarray}
\Rightarrow \frac{1}{2} \; \frac{d L}{d t} &=&
-\int_{V}  \frac{\partial {\bf B}}{\partial t} \cdot [\; w \; \nabla \times ({\bf \Omega_a} \times {\bf B} )]  \; d^3x \nonumber\\
&-& \int_{S} {\bf \hat n} \cdot \left[\; w \; ({\bf \Omega_a} \times {\bf B} ) \times \frac{\partial {\bf B}}{\partial t}  \right]  \; d^2x \nonumber\\
&+& \int_{V}  \left[({\bf \Omega_a} \times {\bf B} ) \times \frac{\partial {\bf B}}{\partial t}
\right] \cdot \nabla w  \; d^3x \nonumber\\
&+& \int_{V} \frac{\partial {\bf B}}{\partial t} \cdot
(\; w \; {\bf \Omega_a} \times (\nabla \times {\bf B})) \;     \; d^3x \nonumber\\
&-&\int_{V} \; w \; \nabla ({\bf \Omega_b} \cdot {\bf B}) \;  \cdot  \frac{\partial {\bf B}}{\partial t}
\; d^3x \nonumber\\
&+&\int_{S} {\bf \hat n}  (\; w \; {\bf \Omega_b} \cdot {\bf B}) \cdot \frac{\partial {\bf B}}{\partial
t} \; d^2x \nonumber\\
&-&\int_{V} \left[ ({\bf \Omega_b} \cdot {\bf B}) \; \nabla w   \right] \;
\cdot \frac{\partial {\bf B}}{\partial t} d^3x \nonumber\\
&+&\int_{V} \; w \; \left[{\bf \Omega_b} \;
(\nabla \cdot {\bf B})\right] \cdot \frac{\partial {\bf B}}{\partial t}  \; d^3x \nonumber\\
&-& \int_{V}  \; w \; (\Omega_a^2 + \Omega_b^2) \; {\bf B} \cdot \frac{\partial {\bf B}}{\partial t} \; d^3x
\end{eqnarray}
Now all terms except the terms 2 and 3 have the correct form. We apply
${\bf a} \cdot ({\bf b} \times {\bf c})={\bf c} \cdot ({\bf a} \times {\bf b})$   to term 2  and 3
\begin{eqnarray}
\Rightarrow \frac{1}{2} \; \frac{d L}{d t} &=&
-\int_{V}  \frac{\partial {\bf B}}{\partial t} \cdot [\; w \; \nabla \times ({\bf \Omega_a} \times {\bf B} )]  \; d^3x \nonumber\\
&-& \int_{S} \left[\; w \; {\bf \hat n} \times ({\bf \Omega_a} \times {\bf B} )\right] \cdot  \frac{\partial {\bf B}}{\partial t}    \; d^2x \nonumber\\
&+& \int_{V} \left[\nabla w \times ({\bf \Omega_a} \times {\bf B} )\right] \cdot  \frac{\partial {\bf B}}{\partial t}    \; d^2x \nonumber\\
&+& \int_{V} \frac{\partial {\bf B}}{\partial t} \cdot
({\bf \Omega_a} \times (\nabla \times {\bf B})) \;     \; d^3x \nonumber\\
&-&\int_{V} \; w \; \nabla ({\bf \Omega_b} \cdot {\bf B}) \;  \cdot  \frac{\partial {\bf B}}{\partial t}
\; d^3x \nonumber\\
&+&\int_{S} {\bf \hat n}  (\; w \; {\bf \Omega_b} \cdot {\bf B}) \cdot \frac{\partial {\bf B}}{\partial
t} \; d^2x \nonumber\\
&-&\int_{V} \left[ ({\bf \Omega_b} \cdot {\bf B}) \; \nabla w   \right] \;
\cdot \frac{\partial {\bf B}}{\partial t} d^3x \nonumber\\
&+&\int_{V} \; w \; \left[{\bf \Omega_b} \;
(\nabla \cdot {\bf B})\right] \cdot \frac{\partial {\bf B}}{\partial t}  \; d^3x \nonumber\\
&-& \int_{V} \; w \; (\Omega_a^2 + \Omega_b^2) \; {\bf B} \cdot \frac{\partial {\bf B}}{\partial t} \; d^3x
\end{eqnarray}
Now all terms have the correct form and we write them more compact
\begin{equation}
\Rightarrow \frac{1}{2} \; \frac{d L}{d t}=-\int_{V} \frac{\partial {\bf B}}{\partial t} \cdot {\bf \tilde{F}} \; d^3x
-\int_{S} \frac{\partial {\bf B}}{\partial t} \cdot {\bf \tilde{G}} \; d^2x
\end{equation}
\begin{eqnarray}
{\bf \tilde{F}}&=& w \; {\bf F} +({\bf \Omega_a} \times {\bf B}
)\times \nabla w  +({\bf \Omega_b} \cdot {\bf B}) \; \nabla w \\
{\bf \tilde{G}}&=& w \; {\bf G}
\end{eqnarray}
\begin{eqnarray}
{\bf F} & =& \nabla \times ({\bf \Omega_a} \times {\bf B} )
- \bf \Omega_a \times (\nabla \times \bf B)  \nonumber\\
& & +\nabla(\bf \Omega_b \cdot \bf B)-  \bf \Omega_b(\nabla \cdot \bf B)
+( \Omega_a^2 + \Omega_b^2)\; \bf B
\end{eqnarray}
\begin{eqnarray}
{\bf G} & = & {\bf \hat n} \times ({\bf \Omega_a} \times {\bf B} )
-{\bf \hat n} (\bf \Omega_b \cdot \bf B)
\end{eqnarray}

\end{article}
\end{document}